# Solutions of nonlinear equation of the curvilinear electromagnetic wave theory for point and non-point electron

Alexander G. Kyriakos

*Saint-Petersburg State Institute of Technology,
St. Petersburg, Russia.*

*Present address: Athens, Greece, e-mail: agkyriak@yahoo.com*

**Abstract**
In previous paper (Kyriakos, 2005) we have shown, that there is a special kind of non-linear electrodynamics - Curvilinear Wave Electrodynamics (CWED), whose equations are mathematically equivalent to the equations of quantum electrodynamics. The purpose of the present paper is to show that in framework of CWED the known solutions of the nonlinear electromagnetic equations can be considered as the approximate solutions of the nonlinear equation of CWED. Another purpose of this paper is to show, that these solutions allow the description of electron-like particle of CWED as point or non-point particles, depending on mathematical approach.

## 1. Introduction

**1.1. Statement of the problem**

The theory of calculation of charge, mass and other characteristics of electron on the basis of the field equations has arisen originally in classical electrodynamics (Kelvin, Larmor, Lorentz, Poincare, etc.). It is based on hypotheses of the field mass and the field charge, according to which, particles' own energy or mass is obliged to energy of fields, and the charge of particles is defined by these particles' own fields. These ideas afterwards were transferred to quantum mechanics. But neither classical, nor quantum theories could explain consistently the nature of mass and charge of elementary particles, and could not deduce numerical values of their charges and masses. The electron is an exception, for which some consecutive theories have been constructed.

**1.2. The general requirements to the classical field mass theory**

The hypothesis of field mass of electron within the framework of classical electrodynamics (Sokolov and Ivanenko, 1949) has been put forward in the end of the $19^{th}$ century by J.J. Thomson and advanced by H. Lorentz, M. Abraham, A. Poincare, etc.

According to this hypothesis the electron's own energy (or its mass) is completely caused by the energy of the electromagnetic field of electron. In the same way it is supposed that the electron momentum is obliged to the momentum of the field. Since electron, as any mechanical particle, possesses the momentum and energy, which are together the 4-vector of the generalized momentum, the necessary condition of success of the theory will be the proof that the generalized momentum of an electromagnetic field is a 4-vector.

Thus, for the success of the field mass theory the following conditions should be satisfied at least:

**At first**, it is necessary to receive final value of the field energy, generated by a particle, which could be precisely equated to final energy of a particle (i.e. product of the mass by the square of the light speed).

**At second**, the value of a momentum of the field, generated by a particle, should not only be final, but also taking place in a proper correlation with energy, forming with the last a four-dimensional vector.

**Thirdly**, the theory should manage to deduce the equation of movement of electron.

**Fourthly**, it is necessary to demand the obtaining of electron spin, as a spin of a field (that needs the quantum generalization of the theory of field mass, since spin is quantum effect).

The analysis shows, that there are two conditions, by which the generalized field momentum $G_\mu$ is a 4-vector.

1) In case of space without charges the size

$$G_\mu = \frac{i}{c} \int T_{\mu 4}(dr), \qquad (1.1)$$

will represent a 4-vector if divergence of energy tensor of a field turns into zero:

$$\frac{\partial T_{\mu\lambda}}{\partial x_\lambda} = 0, \qquad (1.2)$$

For example, the electromagnetic field, which is located in a space without charges, satisfies similar conditions. In particular, due to this fact, in the theory of photons EM field is characterized not only by energy, but also by momentum.

2) The condition, by which the energy and momentum of an electromagnetic field form a 4-vector at the presence of charges, is formulated by the Laue theorem. According to the Laue theorem, at the presence of charges the value $G_\mu$ is a 4-vector only in the case that in the coordinate system, relatively to which electron is in rest, for all the energy tensor components the following parity is observed

$$\int T_{\mu\nu}^0 (d\vec{r}_o) = 0, \qquad (1.3)$$

where $(d\vec{r}_o)$ is elementary volume in reference system, in which the electron is in rest. The equality (1.3) expresses a necessary condition, by which the whole particle charge should be in balance. The component $T_{44}^0$, the integral of which is a constant and is equal to full energy of the field generated by particle, is here excepted. We can equate this field energy to the particle's own energy, expressing in this way the basic idea of a field hypothesis. According to the last:

$$m_e = \frac{\varepsilon_e}{c^2} = \frac{1}{c^2} \int T_{44}^0 (d\vec{r}_o), \tag{1.4}$$

Thus, the mass of a particle from the field point of view can be defined in two ways:
- proceeding from EM momentum of a field $G_1$ it is possible to define mass as factor of proportionality between a field momentum and three-dimensional speed of a particle.
- if we consider the electron's own energy as equal or conterminous to the energy of a field, and mass as the ratio of a field energy $\frac{c}{i}G_4$, to a square of light speed (i.e. as the fourth component of a generalized momentum).

The attempts to execute this program, proceeding from classical linear Maxwell theory, have led to difficulties. In particular, it was not possible to prove the Laue theorem (Tonnelat, 1959). In the classical theory the dynamics (mechanics) and electrodynamics are completely independent from each other. Electromagnetic actions are characterized by component $T_0^0$ of an energy-momentum tensor of an electromagnetic field. It does not include the energy and momentum of the substance, which should be subsequently inserted. The attempts of Lorentz and Poincare to coordinate the theory on the basis of the assumption that energy of substance has an electromagnetic origin, have not led to a positive result. In Lorentz electron theory (linear in essence) existence of charges it is possible to explain only by introduction of forces of non-electromagnetic origin.

Nevertheless (Sokolov and Ivanenko, 1949), there were also a number of successes, which carried a hope to solve this problem by any change of the theory. The most perspective change of Maxwell-Lorentz theory appeared to be its nonlinear generalization.

**1.3. Nonlinear electrodynamics**

Thus, at first, the nonlinear electrodynamics should allow us to calculate a particle's (e.g. of electron) own characteristics– mass, charge, spin etc., proceeding from its own electron field. Besides, the nonlinear field theories essentially differ from the linear theories: in general case they allow us to deduce the equations of movement of particles from the equations of a field, generated by them.

In previous papers (Kyriakos, 2004) within the framework of Curvilinear Waves Electrodynamics (CWED) we have received the nonlinear equation for the electromagnetic (EM) electron-like particle, describing EM structure of such particle, and have shown, that on sufficient distance from a particle it coincides with the linear Dirac electron equation. Thus, such EM particle possesses the charge and spin of electron. At the same time the solution of the nonlinear equation of the curvilinear electromagnetic wave in a general case is not received yet. Its first approach – the nonlinear Heisenberg equation - also did not manage fully to be solved, although here the encouraging results have been received.

We will show, that in framework of CWED the known approaches of Gustav Mie, M.Born - L. Infeld, E. Schroedinger etc. represent the approximate solutions of nonlinear equation of CWED for an EM electron-like particle. These solutions enable us to estimate the sizes of a particle and distribution of a field in approach of spherical electron.

Besides, the nonlinear theories find out an opportunity of description of EM electron-like particles as point or not point, depending on the used mathematics.

## 2.0. Gustav Mie approach

**2.1. Prior history**

Gustav Mie made the first attempt to construct a purely nonlinear electromagnetic theory of charged particles (Mie, 1912a; 1912b; 1913), (Pauli, 1958), (Tonnelat, 1959). Proceeding from some formally irreproachable hypothetical nonlinear generalization of electrodynamics, he managed to construct a theory, which has overcome all difficulties of the classical theory.

As we have said above, in the theory of the electron before Mie (Bialynicki-Birula, 1983), the electron was not treated as a purely electromagnetic entity, but it was also made of other stuff, like, for example, Poincare stresses and the mechanical mass. Mie wanted only the electromagnetic field to be responsible for all the properties of the electron. In particular, he wanted the electromagnetic current to be made of electromagnetism. In order to achieve this goal, Mie assumed that the potential four-vector enters directly into the Lagrangian and not only through the field strength.

The generation of the current has been achieved in this manner, but the price was very high. The potentials acquired a physical meaning and the gauge invariance was lost. This property has been found unacceptable by other physicists and the theory of Mie has been shelved for two decades.

**2.2. G. Mie theory**

In his theory Mie has made two essential steps (Pauli, 1958), (Tonnelat, 1959). At first, Mie was the first who suggested leaning in the construction of the theory on a Lagrangian, dependent on fundamental invariants. At second, to get rid of Poincare–Lorentz, forces that have non-electromagnetic origin, Mie entered a uniform sight at a field and substance. He set a problem in order to generalize the equations of a field and an energy-momentum tensor of Maxwell-Lorentz theory in a way that inside the elementary charged particles the repulsion Coulomb forces were counterbalanced by other forces, also with electric origin, and outside of particles the deviation from ordinary electrodynamics was imperceptible. He assumed that any energy and substance has an electromagnetic origin, and sets as purpose to deduce properties and characteristics of charges, proceeding from properties of a field

About the kind of Lagrangian *L*, which is frequently called a world function, in nonlinear electrodynamics it is possible to make some general statements. The independent invariants, which can be formed from EM bivector $F_{\mu\nu}$ (where $F_{\mu\nu}$ are the tensor components of electromagnetic field strengths) and a vector $A_\mu = (i\varphi, \vec{A}) = (i\varphi, A_i) = (A_4, A_i)$ of an electromagnetic field are the following:

1. The square of bivector $F_{\mu\nu}$: $I_1 = \frac{1}{4} F_{\mu\nu} F^{\mu\nu}$; 2. the square of a pseudo-vector $I_2 = \frac{1}{4} F_{\mu\nu} F^{*\mu\nu}$ (where $F^{*\mu\nu}$ is the dual electromagnetic tensor). 3. The square of a 4-vector of electromagnetic potential $A_\mu$: $I_3 = A_\mu A^\mu$; 4. The square of a vector: $I_4 = F_{\mu\nu} A_\mu$; 5. The square of a vector: $I_5 = F^*_{\mu\nu} A_\mu$.

Therefore $L$ can depend only on these five invariants. If $L$ is equal to the first of the specified invariants, the field equations are degenerated into ordinary equations of the electron theory for space without charges. Thus, $L$ can noticeably differ from $\frac{1}{4} F_{\mu\nu} F^{\mu\nu}$ only inside the material particles. Invariant 2 can be included in $L$ only as a square, in order not to break the invariance, concerning spatial reflections. Invariants 3-5 break the gauge invariance. Further statements about the world function $L$ cannot be made. Thus, for the selection of $L$ there are an infinite number of opportunities.

Gustav Mie accepted as initial the following Lagrangian:

$$L_{Mi} = \frac{1}{4} F_{\mu\nu} F^{\mu\nu} - f\left(\pm \sqrt{A_\mu A^\mu}\right), \qquad (2.1')$$

or

$$L_{Mi} = \frac{1}{8\pi}\left(E^2 - H^2\right) - f\left(\pm \sqrt{A_\mu A^\mu}\right), \qquad (2.1'')$$

where $f$ is some function, $\vec{E}$, $\vec{H}$ are the vectors of electric and magnetic field strengths, accordingly.

Using this Lagrangian (Tonnelat, 1959), Gustav Mie managed to receive the final energy (or mass) of the charged particle as a value completely caused by the energy of the field of this particle. Besides, in this theory the Laue theorem of stability is carried out and the proper correlation between energy and momentum of a particle is reached.

For further analysis it is also useful to mention the attempt of H. Weyl (Pauli, 1958) to interpret on the basis of Mie theory the asymmetry (distinction of masses) of both sorts of electricity. If $L$ is not a rational function of $\sqrt{A_\mu A^\mu}$, it is possible to put:

$$L^+_{Mi} = \frac{1}{4} F_{\mu\nu} F^{\mu\nu} - f\left(+\sqrt{A_\mu A^\mu}\right), \qquad (2.2')$$

$$L^-_{Mi} = \frac{1}{4} F_{\mu\nu} F^{\mu\nu} - f\left(-\sqrt{A_\mu A^\mu}\right), \qquad (2.2'')$$

Thus, if $L$ is a multiple-valued function of the invariants mentioned above, it is obviously possible to choose as world functions for positive and negative charges various unequivocal branches of this function.

### 2.3. Connection of the Mie theory with CWED

Let's show that the Mie Lagrangian after some additions can be submitted as Lagrangian similar to Lagrangian of CWED (and consequently, of QED).

As we know (Pauli, 1958; Sommerfeld, 1958), the charge density is not invariant concerning Lorentz transformation, but a charge is absolute invariant concerning this.

Also it is known, that the square of 4-potential, i.e. $I_3 = A_\mu A^\mu$, is invariant concerning Lorentz transformation, but it is not invariant relatively to gauge transformations.

But it appears, that the product of a square of a charge on $I_3$ will be an invariant concerning both Lorentz and gauge transformations.

### 2.3.1. Larmor - Schwarzschild invariant

According to (Pauli, 1958; Sommerfeld, 1958), K. Schwarzschild (Schwarzschild, 1903) entered the value

$$S_w = \varphi - \frac{\vec{\upsilon}}{c} \cdot \vec{A}, \qquad (2.3)$$

which he called "electrokinetic's potential", and has shown, that this value, being multiplied by density of a charge $\rho$, forms the relativistic invariant:

$$L' = \rho(\varphi - \frac{\vec{\upsilon}}{c} \cdot \vec{A}) = -\frac{1}{c} j_\mu \cdot A^\mu = \rho S_w, \qquad (2.4)$$

where $j_\mu = \{ic\rho, \rho\vec{\upsilon}\}$ is 4-current density, $A^\mu = \{i\varphi, \vec{A}\}$ is 4-potential. Schwarzschild forms the Lagrange function by integration on space

$$L = \frac{1}{2} \int (H^2 - E^2) dV + \int \rho(\varphi - \frac{\vec{\upsilon}}{c} \cdot \vec{A}) dV, \qquad (2.5)$$

then he receives the function of action by integration on time.

Thus, in 4-dimensional designations the Lagrange function density (or Lagrangian) will be written down as follows:

$$\bar{L} = \frac{1}{4} F_{\mu\nu} F^{\mu\nu} - \frac{1}{c} j_\mu A^\mu, \qquad (2.6)$$

and the Lagrange function will be:

$$L = \frac{1}{4} \int F_{\mu\nu} F^{\mu\nu} d\tau - \frac{1}{c} \int j_\mu A^\mu d\tau, \qquad (2.7)$$

(In the note 10 to the book (Pauli, 1958) Pauli marked, that before Schwarzschild the same Lagrangian has been suggested by J.J. Larmor (Larmor, 1900)).

We will consider now the function:

$$A_\mu^2 \equiv A_\mu A_\mu = -\varphi^2 + A_i^2, \qquad (2.8)$$

which enters in Lagrangiane of Mie. Multiplying it on the squares of density of a charge and a square of a charge, we shall receive accordingly:

$$e^2 A_\mu^2 = -(e\varphi)^2 + (e\vec{A})^2, \qquad (2.9)$$

We will enter the values of density of energy of interaction and energy of electron interaction, accordingly:

$$U_e = \rho\varphi, \; \varepsilon_e = e\varphi, \qquad (2.10)$$

and also the density of momentum and the momentum of electron interaction, accordingly:

$$g_{ei} = \frac{1}{c}\rho A_i, \quad p_{ei} = \frac{1}{c}eA_i, \tag{2.11}$$

Then from (2.9) we will receive:

$$e^2 A_\mu^2 = -\varepsilon_e^2 + (c\vec{p}_e)^2, \tag{2.12}$$

As $(\hat{\alpha}_o \varepsilon)^2 = \varepsilon^2$, $(\hat{\vec{\alpha}}\ \vec{p})^2 = \vec{p}^2$ these expressions can be also written down as:

$$e^2 A_\mu^2 = -(\varepsilon_e^2 - c^2 p_{ei}^2) = -\left((\hat{\alpha}_0 \varepsilon_e)^2 - c^2(\hat{\vec{\alpha}}\vec{p}_{e_i})^2\right), \tag{2.13}$$

Using the above-stated results, for nonlinear part of Mie Lagrangian $L_{Mie}^N = f\left(\pm \sqrt{A_\mu A^\mu}\right)$, we will accept the expression:

$$L_{Mie}^N = \rho\left(\pm \sqrt{\varphi^2 - c^2 \vec{A}^2}\right), \tag{2.14}$$

it is easy to receive, using properties of Dirac matrixes, the following decomposition:

$$\sqrt{e^2 A_\mu^2} = \mp(\hat{\alpha}_0 \varepsilon_e \pm c\hat{\vec{\alpha}}\vec{p}_e), \tag{2.15}$$

That gives for Lagrangian the expression:

$$L_{Mie}^{Ne} = \mp(\hat{\alpha}_0 \varepsilon_e \pm c\hat{\vec{\alpha}}\vec{p}_e), \tag{2.16}$$

Taking into account that

$$\hat{\beta}\ mc^2 = -(\varepsilon_p - c\hat{\vec{\alpha}} \cdot \vec{p}_p), \tag{2.17}$$

we see that we can enter in the Lagrangian the mass term of the Dirac equation. Thus, it is possible to assert, that Mie Lagrangian can be transformed so, in order to have the form of the Lagrangian of the nonlinear field theory, corresponding to the theory of EM electron-like particles (CWED).

The use of these expressions leads to the Dirac equations of electron and positron, and offers the basis to H. Weyl's attempt to interpret the asymmetry of both sorts of electricity, not in connection with mass, but in connection with distinction particle - antiparticle.

Also it is easy to see connection of the Mie Lagrangian with Born - Infeld Lagrangian. Actually taking into account that $e^2 A_\mu^2 = -\left[\left(\int_{-\infty}^{+\infty} U_e^2 dV\right)^2 - c^2\left(\int_{-\infty}^{+\infty} \vec{g}_e^2 dV\right)^2\right]$

and remembering the electromagnetic representation of Fierz indentity (Kyriakos, 2004):

$$(8\pi)^2(U^2 - c^2\vec{g}^2) = (\vec{E}^2 + \vec{H}^2) - 4[\vec{E}\times\vec{H}]^2 = (\vec{E}^2 - \vec{H}^2) - 4(\vec{E}\cdot\vec{H})^2, \tag{2.18}$$

we can obtain the same term as in the Born-Infeld Lagrangian (see below).

Thus, the assumption of Mie that internal properties of electron are described by an electromagnetic field, corresponds to the results of CWED for EM electron-like particles. Actually, in framework of CWED we came to the conclusion (Kyriakos, 2004), that the product of potentials with charge in Dirac equation are equivalent to internal energy of a particle, since through them the mass of a particle or an internal electric current of a particle is expressed. If to accept, that potentials inside a particle correspond to an energy-momentum of a field of the particle, it makes the potentials physically certain values, which however are not measurable outside of a particle. In other words, the potentials are here the hidden parameters of EM elementary particles.

Do these results contradict to the experimental results of modern physics?

As it is known in classical electrodynamics the potentials play the role of the mathematical auxiliary values and have no physical sense. But as it appears, in framework of quantum mechanics the potentials have physical sense, that is proved by Aharonov-Bohm experiment (Aharonov and Bohm, 1959), (see discussion in (Feynman, Leighton and Sands, 1989)).

As an example of approximate calculation of parameters of an EM electron-like particle in framework of CWED, we will consider the results of the Born - Infeld theory (Born and Infeld, 1934).

## 3.0. Born-Infeld nonlinear theory

Born and Infeld revived Mie's theory and proposed a specific model. The Born-Infeld theory is derived from the simplest possible Lagrangian: the square root of the determinant of a second rank covariant tensor. Such a structure automatically guarantees the invariance of the theory under arbitrary coordinate transformations, making the fully relativistic and gauge invariant nonlinear electrodynamics.

**3.1. The Born-Infeld nonlinear theory results**

M. Born and L. Infeld proceeded (Born and Infeld, 1934) from the idea of a limited value of the electromagnetic field strength of the electron. This reason and some others led them to the following Lagrangian of the nonlinear electrodynamics in the vacuum:

$$L_{BI} = \frac{E_0^2}{4\pi}\left(1 - \sqrt{1 - \frac{E^2 - H^2}{E_0^2} - \frac{(\vec{E}\cdot\vec{H})^2}{E_0^4}}\right), \quad (3.1)$$

where $E_0$ is the maximum field of electron.

In short the results of the Born - Infeld theory for the most simple case of an electrostatic field - spherical electron - are following. For an electrical induction we obtain here:

$$D_r = \frac{e\vec{r}}{r^3}, \quad (3.2)$$

*As we see, from point of view of the D-field, the electron should be considered as point particle.*

For the electric field (E-field) we obtain:

$$\vec{E}_r = \frac{\vec{D}_r}{\sqrt{1 + \frac{D_r^2}{E_0^2}}} = \frac{e\vec{r}}{r\sqrt{r^4 + r_0^4}}, \quad (3.3)$$

where $r_0 = \sqrt{\frac{e}{E_0}}$. In this case, i.e. *from point of view of the electric field (E-field), the electron is not a point particle.*

This is very important specificity of the nonlinear theory in comparison with the linear theory, which can explain, why experiments on electron-photon or electron-electron scatterings can be interpreted so, that the electron looks as a point particle.

Thus, from the point of view of the analysis of an electric field the electron is not point, and its charge density distribution can be found:

$$\rho = \frac{divE}{4\pi} = \frac{er_0^4}{2\pi \, r(r^4 + r_0^4)^{3/2}}, \qquad (3.4)$$

Differently, from the point of view of calculation of an electric field the charge can be considered as distributed (mainly) in volume of radius $r_0$ since by $r \gg r_0$ the density will quickly aspire to zero. Therefore the size $r_0$ can be considered as effective radius of electron.

Using known values for mass and charge of electron and speed of light, it is possible to receive here the effective radius of electron, which is practically equal to classical radius of electron: $r_0 = 2{,}28 \cdot 10^{-13}$ см.

Also it is easy to find value for the maximal field of electron, being a field in the center of the electron (at $r = 0$): $E_0 = \frac{e}{r_0^2} = 9{,}18 \cdot 10^{15} \, CGS = 2{,}75 \cdot 10^{20} \, \frac{V}{m}$.

As it is known (Ivanenko and Sokolov, 1949), the two types of fields and the two definitions of the charge density, corresponding to them, are also described by the theory of the dielectrics. The value:

$$\varepsilon = \frac{D}{E} = \sqrt{\frac{r^4 + r_0^4}{r^4}}, \qquad (3.5)$$

which is here a function of the position, can be considered as a "dielectric permeability of electron". On large distances from a charge, when $\frac{r_0}{r} \to 0$, $\varepsilon$ acquires a value equal to unit as in usual electrodynamics. It is possible to tell, that instead of the expression of energy $\frac{e^2}{r^2}$, Born and Infeld take $\frac{e^2}{\varepsilon \, r^2}$, and then the reduction of $r$ is compensated by increase of $\varepsilon$ so the full energy remains as final. (It is possible to assume, that the presence of physical vacuum should make the amendment to value of dielectric permeability, and at the same time, in values of potential of electron, its size and other characteristics).

Thus (Ivanenko and Sokolov, 1949), proceeding from some formal hypothetical nonlinear generalization of electrodynamics, it appeared possible:
1. to prove the theorem of stability, i.e. to prove, that in the nonlinear theory the electron is stable without introduction of forces of non-electromagnetic origin;
2. to receive the final energy (mass) of a particle;
3. to receive the final size of an electric charge;
4. to receive the final size of an electromagnetic field.

It is interesting to note that M. Born (Born, 1953) assume that the even more important achievement of this theory represents the estimation of fine structure constant, received by Heisenberg and his collaborators (Euler, 1936; Euler and Heisenberg, 1936; Euler and Kockel, 1935; Kockel, 1937) and confirmed by Weisskopf (Weisskopf, 1936).

It has been done by the comparison of the lowest nonlinear terms of the given theory with the corresponding expressions from the Dirac holes theory, caused what is refered to as "polarization of vacuum". The result is: $\alpha = \frac{e^2}{\hbar c} = 0{,}0122 \left(\frac{1}{\alpha} = 82\right)$ (the estimation of the fine structure constant can be also made in frameworks of CWED by more simple way (Kyriakos, 2002)

### 3.2. Other Lagrangians of nonlinear theories

Also others Lagrangians have been offered for reception of the nonlinear theory. So Schroedinger used the following arbitrary combination for Lagrangian:

$$L_{Sch} = \frac{E_0^2}{8\pi} \ln\left(1 + \frac{E^2 - H^2}{E_0^2}\right), \qquad (3.6)$$

It was noted (Ivanenko and Sokolov, 1949) that various and, as was outlined, from the physical point of view, arbitrary variants of formal nonlinear electrodynamics lead to close values of coefficients, if to take into account, that the electron radius is equal to classical radius of electron.

It was also noted, that the basic defect of these theories, as well as of Mie theory, was the arbitrary choice of Lagrangian, which had no connection with the quantum theory, in particular, with Dirac theory, and did not take into account properties of electron, revealed by the last.

We will show that the results of these theories can be considered as approximation of the CWED theory results for EM electron-like particle, and that they are mathematically connected to the Dirac electron theory.

## 4.0. Connection of the Born-Infeld theory with CWED

As we have shown (Kyriakos, 2004a) in general case the CWED is the nonlinear theory. The Lagrangian of the non-linear field theory can be written generally as some function of the field invariants:

$$\overline{L} = f_L(I_1, I_2), \qquad (4.1)$$

where $I_1 = \left(\vec{E}^2 - \vec{H}^2\right), I_2 = \left(\vec{E} \cdot \vec{H}\right)$ are the invariants of electromagnetic field theory.

Apparently, for each problem the function $f_L$ has its special form, which is unknown. It can suppose that there is an expansion of the function $f_L$ in Taylor – MacLaurent power series with unknown expansion coefficients (Kyriakos, 2004a).

Obviously, for the most types of the functions $f_L(I_1, I_2)$ the expansion contains approximately the same set of the terms, which distinguish only by the constant coefficients, any of which can be equal to zero (as an example of the expansion it is possible to point out the expansion of the quantum electrodynamics Lagrangian for particle at the present of physical vacuum (Akhiezer and Berestetskii, 1965), (Weisskopf, 1936), (Schwinger, 1951)). Generally the expansion looks like:

$$L_M = \frac{1}{8\pi}\left(\vec{E}^2 - \vec{B}^2\right) + L', \tag{4.2}$$

where

$$\begin{aligned}L' = &\alpha\left(\vec{E}^2 - \vec{B}^2\right)^2 + \beta\left(\vec{E}\cdot\vec{B}\right)^2 + \gamma\left(\vec{E}^2 - \vec{B}^2\right)\left(\vec{E}\cdot\vec{B}\right) + \\ &+ \xi\left(\vec{E}^2 - \vec{B}^2\right)^3 + \zeta\left(\vec{E}^2 - \vec{B}^2\right)\left(\vec{E}\cdot\vec{B}\right)^2 + ...\end{aligned} \tag{4.3}$$

is the part which is responsible for the non-linear interaction (here $\alpha, \beta, \gamma, \xi, \zeta,...$ are some constants)

In the case of the Born-Infeld nonlinear electrodynamics the Lagrangian can be expanded into the small parameters $a^2 E^2 \ll 1$ and $a^2 B^2 \ll 1$, where $a^2 = \frac{1}{E_0^2}$, so that we have:

$$L_{BI} = -\frac{1}{8\pi}\left(\vec{E}^2 - \vec{B}^2\right) + \frac{a^2}{32\pi}\left[\left(\vec{E}^2 - \vec{B}^2\right)^2 + 4\left(\vec{E}\cdot\vec{B}\right)^2\right] + \sum O(\vec{E}^2, \vec{H}^2), \tag{4.4}$$

where $\sum O(\vec{E}^2, \vec{H}^2)$ is the series remainder with the terms, containing vectors of the electromagnetic field in powers, higher than the fourth. Obviously, under conditions $a^2 E^2 \ll 1$ and $a^2 B^2 \ll 1$ on large distance from the center of a particle (where there is a maximal field) the terms of these series really quickly converge, but on small distance from the center it is, apparently, not so and demands the account of members of higher and higher degrees.

In the paper (Kyriakos, 2004) we have shown, that at the first approximation Lagrangian of CWED in electromagnetic form can be represented as following:

$$L_N = -\frac{1}{8\pi}\left(\vec{E}^2 - \vec{B}^2\right) + b\left[\left(\vec{E}^2 - \vec{B}^2\right) + 4\left(\left(\vec{E}\cdot\vec{B}\right)^2\right)\right], \tag{4.5}$$

where b is some constant. Taking into account (4.4), we can write

$$L_N \approx L_{BI}, \tag{4.6}$$

and receive in the framework of CWED for EM electron-like particle the approximate solution, like the solution of Born - Infeld theory, stated briefly above.

We can similarly show, that Lagrangian of the nonlinear equation of EM electron-like particle in CWED approximately coincides with Lagrangian of Schroedinger and others offered Lagrangians of nonlinear theories, allowing us to calculate the corresponding characteristics of electron.

Thus, it is not difficult to understand, why "various, from the physical point of view, variants of formal nonlinear electrodynamics lead to close values of coefficients": as expansion of nonlinear Lagrangian (4.3) shows, all of them are approximately equal among themselves and consequently yield close results.

At the same time, since Lagrangian and equations of CWED completely coincide with Lagrangian and the equations of quantum electrodynamics, the Mie theory and its variant – the Born - Infeld theory, is closely connected with the quantum theory, in particular, with Dirac theory.

## 5.0. Discussion about pointness of EM electron-like particles

It seems that the non-point solution for the EM electron-like particle in framework of CWED contradicts to the results of experiments by measuring the electron size. We will analyze below this problem and show that contradiction does not exists.

The fact that the electron sizes are not included into Dirac equation, gives the basis to postulate (Landau and Lifshits, 1962), that electron is a point particle, i.e. a particle, about which we assume, that its mechanical condition is completely described by the task of three coordinates and three component of speed of movement as whole. Such postulate leads to occurrence of infinities of the same type, which we had in the classical electron theory due to Coulomb law. For liquidation of infinity of energy (mass) and charge, arising in this case, a special artificial operation, named renormalization, which allows us to calculate all necessary characteristics with big accuracy, is entered. Roughly the operation of renormalization consists of the replacement of infinite sizes of a charge and mass (energy) of electron with their experimental values.

It is known (Naumov, 1982), that when on the basis of experiments it is mentioned that electron is point particle, actually this means, that quantum electrodynamics is fair for any distances (until today it is checked up to distance $2.10^{-18}$m). Thus, the experiment checks the existing calculation method of QED, not the validity of the concept of pointness of electron. On the contrary in the framework of CWED these experiments can be interpreted as conformation of the electromagnetic origin of particles. Actually since the EM particles of CWED are the specially configured electromagnetic field, the theory must be right in any point of a particle, so that in this sense a particle has not got any size.

As we have shown above, on one hand the approximate solution of nonlinear equation of CWED can be described in way that characterizes a particle, as point. Thus in this case the solution does not conflict with QED, but needs some renormalization as in QED.

On the other hand in framework of CWED (Kyriakos, 2004) the Dirac equation describes an EM particle with certain sizes and form. In representation of currents it includes the radius of "bare" rest electron, which, if we are not taking into account the polarization of physical vacuum, is equal to Compton wavelength of electron.

Such representation is proved by the analysis of the Dirac equation in QED. It is known that (Messiah, 1973) "in the non-relativistic limit, the electron does not represent a point charge, but the distribution of charge and current in the area with linear sizes $\frac{\hbar}{mc}$. This explains the occurrence of interaction terms, which are connected to the magnetic moment (interaction $\vec{\mu}\vec{H}$, spin-orbital interaction) and the distributed density of charges (Darvin component)".

It is not difficult to show (Kyriakos, 2002), that the polarization of physical vacuum leads to screening of electron charge and to reduction of radius of a particle up to size of classical electron radius $r_o = \frac{e^2}{mc^2}$. This too does not contradict to the results of QED and the experiments. Actually, according to calculations by help of the perturbation theory all formulae of computation of QED effects contain the interaction cross-sections, defined by classical electron radius.

Thus the CWED results contradict neither the results of QED nor experimental results.